\documentstyle[aps,twocolumn,epsf]{revtex} 

\begin{document} 
\draft

\twocolumn[\hsize\textwidth\columnwidth\hsize\csname
@twocolumnfalse\endcsname

\tightenlines

\title{The effect of geometrical confinement on the interaction between
 charged colloidal suspensions}

\author{E.Allahyarov
 $^{1,2}$~\cite{kendim},\,\, I.D'Amico $^{3}$,\, H.L\"owen $^{1}$\,\, }

\address{{1} Institut\, f\"{u}r\, Theoretische \,Physik
  II,\,Heinrich-Heine-Universit\"{a}t\,
 D\"{u}sseldorf,\,\mbox{D-40225}\,D\"{u}sseldorf, \,Germany}
\address{{2} Institute\, for\, High\, Temperatures,\, Russian \,Academy \,of \,Sciences,\, 127412
\,Moscow,\, Russia }
\address{{3} Department of Physics and Astronomy, University of 
Missouri-Columbia, Columbia, Missouri 65211, USA}

\date{\today}

\maketitle

\begin{abstract}
The effective interaction between charged colloidal particles 
confined between two planar like-charged walls is investigated using computer
simulations of the primitive model describing  asymmetric electrolytes. In detail,
we calculate the effective force acting onto a single macroion 
and onto a macroion pair in the presence of slit-like confinement. For moderate
Coulomb coupling, we  find that this force is repulsive.
Under strong coupling conditions, however, the sign of the force 
depends on the distance to the plates and on the interparticle distance.
In particular, the particle-plate interaction becomes strongly
attractive for small distances 
which may explain the occurrence of colloidal crystalline layers
near the plates observed in recent experiments.
\end{abstract}

\vskip2pc]

\renewcommand{\thepage}{\hskip 8.9cm \arabic{page} \hfill Typeset
using REV\TeX }
\narrowtext

\section{Introduction}
There is recent experimental evidence that the effective interaction
between like-charged colloidal particles (``macroions") is  sensitively
affected by a confinement between two parallel charged glass plates 
\cite{kepler,crocker,nature}.
For aqueous polystyrene suspensions studied in experiment, the effective force
between two colloidal macroions is found to be repulsive far away from the plates
but becomes attractive when the like-charge macroions are located  close to an equally charged
plate.
At  first glance, these findings are surprising as one would expect a purely
repulsive interaction from the electrostatic part of the
traditional Derjaguin-Landau-Verwey-Overbeek (DLVO) theory \cite{DLVO}.
 In fact a full theoretical
explanation is still missing but several steps were performed in
different directions: the essential difference in a confining geometry with respect to the bulk
is that the counterion
density field is inhomogeneous for small coupling between the macroions and counterions.
In a straightforward generalization of the DLVO theory to such an inhomogeneous
situation \cite{anne,Jay}, the effective force between the macroions 
remains repulsive close to the charged
plates but  becomes weaker since the local concentration of counterions is higher
which results in a stronger screening of the Coulomb repulsion. It was further realized
that a charged wall induces significant effective triplet interactions \cite{HLEA}
which are
ignored in the usual DLVO approach resulting in a net attraction \cite{dave23}
or in a repulsion \cite{Tehver} depending on the system parameters.
An explicit calculation was done within density functional perturbation theory
which is justified, however, only  for weak inhomogeneities. 
A complementary approach is to solve the nonlinear Poisson-Boltzmann equation with appropriate
boundary conditions in a finite geometry. This was done recently for two charged spheres in a charged
cylindrical pore \cite{bowen} as well as for two charged
cylinders confined by two parallel charged plates \cite{ospeck}. However, the first situation
corresponds to a finite system where the Poisson-Boltzmann approach does not lead to attraction
\cite{Neu} and the second situation is a quasi-two-dimensional set-up which is known to behave
qualitatively different from  a three-dimensional situation \cite{Schmitz}.
A further complication
arises from image charges induced by the different dielectric constants of the glass
and the solvent \cite{Chang,Tandon,dave1}.

A general problem of any theoretical description (as DLVO, Poisson-Boltzmann)
is that close to the walls
the counterion concentration is high and any weak-coupling theory fails {\it a priori\/}
when applied to a situation of confined macroions. For strong coupling, even in the
bulk, it is unclear whether an effective attraction of like-charged spherical macroions
is possible although there are hints from experiments \cite{Ise,Horn,Weiss}, theory \cite{Levin,Netz,shklov,Preparata,Tokuyama},
and computer simulations \cite{allah,Pincus,Tang}. At this stage it is important to remark
that a phase separation seen in experiment does not necessarily imply an effective
attraction. The additional contribution from the counterions to the total free energy
may induce such a phase separation although the effective interaction between the macroions
is purely repulsive \cite{roij,Graf}. Bearing the difficulties in experimental interpretations and 
theory in mind, computer simulations represent a helpful alternative 
tool to extract ``exact" results for certain model systems. The general accepted theoretical
model for the description of charged colloidal suspensions is the ``primitive approach" where
the discrete structure of the solvent is disregarded and the interaction between the macroions
and counterions is modelled by excluded volume and Coulomb forces. The problem with a full
computer simulation of the primitive model is the high charge asymmetry between macro- and counterions
which restricts the full treatment to micelles rather than charged colloidal suspensions \cite{Linse}.

In this paper, we use computer simulations to obtain ``exact" results for the effective interaction
between confined charged colloids based upon the primitive model. Instead of solving the full
many-body problem with many macroions, we only simulate one or two macroions confined between
two parallel charged plates. This enables us to access high charge numbers of the macro-particles.
As a result we find that the wall-particle and the interparticle interaction is repulsive for
weak Coulomb coupling. For stronger coupling, the behaviour of the force changes from
repulsive to attractive and back to repulsive 
as the interparticle distance is varied.  In particular, the
plate-particle interaction exhibits a short-range attraction for a
small distances. This may explain the occurrence
of crystalline colloidal layers on top of the glass plates found in recent experiments
\cite{Larsen_crystallites,larsen,jessica}. These crystallites are metastable but very long-lived 
and cannot be understood in terms of DLVO-theory.

The paper is organized as follows: the model and our target quantities
are defined in section II\@. Section III contains details of our
computer simulation procedure. Results for the counterion density
profiles are shown in section IV\@. The case of a single macroion is discussed
in section V, and a macroion pair is investigated in section VI\@.
Finally, we conclude in section VII\@.

\section{The model and target quantities}

We consider $N_m$ macroions with bare charge $q_{m}=Ze>0$ ($e>0$ denoting the elementary charge)
and mesoscopic diameter $d_m$ confined between two parallel 
 plates that carry surface charge densities $\sigma_{1}$ and
$\sigma_{2}$. We assume that the plates and the macroions are likely charged.
The separation distance between plates is $2L$.  For convenience,
we choose the $z$ axis to be perpendicular to the plate surface.  The origin of the coordinate
system is located on the surface of one plate. Image charges are neglected, i.e.\
we assumed for simplicity that the dielectric constants of the solvent, the plate and the colloidal 
material are the same.
Typically we use a periodically repeated square cell in $x$ and $y$ direction
which possesses an area $S_{p}$. Hence the macroion number density is $\rho_{m}=N_m/2LS_p$.
We restrict our studies to a small number of macroions in the cell. In particular we are considering
the cases $N_m=0,1,2$ subsequently. Both the macroions and the charged plates provide
neutralizing counterions which are dissolved in a solvent of dielectric constant $\epsilon$.
The  counterions have a microscopic diameter $d_c$ and 
carry an opposite charge $q_c = -qe$  where $q>0$ denotes the valency.
Typically, $q=1,2$. For
 simplicity, we assume that the counterions from the walls and from the macroions 
 are not distinguishable. The total counterion number $N_c$ in the cell 
(as well as the averaged counterion number density $\rho_{c}=N_c/2LS_p$)
 is fixed by the condition of global charge neutrality
\begin{equation}
\label{neutral1}
\rho_{m}q_{m}+\rho_{c}q_{c}+\frac{\sigma_{1}+\sigma_{2}}{2L} = 0.
\end{equation}

The interactions between the particles  are described within the framework
of the primitive model. We assume the following pair interaction potentials $V_{mm}(r)$,
 $V_{mc}(r)$, $V_{cc}(r)$ between
macroions and counterions, $r$ denoting the corresponding interparticle distance:
\begin{equation}
V_{mm} (r) = \cases {\infty &for $ r \leq d_m$\cr
   {{Z^2e^2} \over {\epsilon r}} &for $ r > d_m$\cr}
\label{1aLMH}
\end{equation}

\begin{equation}
V_{mc} (r) = \cases {\infty &for $ r \leq (d_m + d_c) /2$\cr
  - {{Zqe^2} \over {\epsilon r}} &for $ r > (d_m + d_c) /2$\cr}
\label{1bLMH}
\end{equation}

\begin{equation}
V_{cc} (r) =\cases {\infty &for $ r \leq d_c$\cr
   {{q^2e^2} \over {\epsilon r}} &for $ r > d_c$\cr}
\label{1cLMH}
\end{equation}
The interaction between the particles and the wall is described by the potential energy
\begin{equation}
 V_{pi} (z) =\cases {\infty &for $z<d_i/2$ and \cr
 & $z>2L-d_i/2$\cr
   {{2\pi(\sigma_2-\sigma_1)q_i z} \over {\epsilon }}&else \cr}
\label{2} 
\end{equation}
where $z$ is the altitude of the particle center and $i=m,c$. Note that the interaction between the wall and the particles is zero
for equally charged plates.

Our target quantities are the equilibrium counterion profiles and the effective forces
exerted on the macroions. The counterionic density profile
$\rho_c^{(0)}(\vec r)$ is 
defined as statistical
average via
\begin{equation}
\rho_c^{(0)}({\vec r})= \sum_{j=1}^{N_c}<\delta ( {\vec r} - {\vec
 r}_j )>_c
\label{3}
\end{equation}
where $\{ {\vec r}_j=(x_j, y_j, z_j); j=1,...N_c \}$ denote the counterion positions.
 The canonical
average  $<...>_c$  over an $\{ {\vec r}_j \}$-dependent
quantity $\cal A$ is defined via the classical trace
\begin{eqnarray}
<{\cal A}(\{ {\vec r}_k \}) >_c =&& {1\over {\cal Z}} {1\over {N_c!}}
 \int_V d^3r_1...\int_V d^3r_{N_c}\nonumber \\ 
&&\times {\cal A} (\{ {\vec r}_k \})
 \exp \left( - {{V_c}\over{k_BT}} \right)
\label{33}
\end{eqnarray}
where $k_BT$ is the thermal energy ($k_B$ denoting Boltzmann's constant) and
\begin{eqnarray}
V_c =&&\sum_{n=1}^{N_m} \sum_{j=1}^{N_c} V_{mc}( \mid {\vec R}_n - {\vec r}_j \mid )
\nonumber \\ 
& & + {{1} \over {2}} \sum_{i,j=1; i\not= j}^{N_c} V_{cc}( \mid {\vec r}_i
- {\vec r}_j \mid ) +
\sum_{j=1}^{N_c} V_{pc}(z_j) 
\label{9999}
\end{eqnarray}
is the total counterionic part of the potential energy provided the
macroions are at positions\\ \hbox {$\{ {\vec R}_j=(X_j, Y_j, Z_j); j=1,...N_m \}$}. Furthermore, the classical
partition function 
\begin{equation}
{\cal Z} = {1\over {N_c!}}
 \int_V d^3r_1...\int_V d^3r_{N_c}
\exp \left( - {{V_c}\over{k_BT}}\right)
\label{neu3}
\end{equation} 
guarantees the correct normalization $<1>_c=1$. Note that the counterionic density profile
$\rho_c^{(0)}({\vec r})$ depends parametrically on the macroion positions 
$\{ {\vec R}_j \}$.

The total effective force ${\vec F}_j$ acting onto the $j$th macroion 
contains three different parts \cite{LMH,allah2,allah}
\begin{equation}
{\vec F}_j = {\vec F}_j^{(1)} + {\vec F}_j^{(2)} + {\vec F}_j^{(3)} 
\label{neu4}
\end{equation} 
The first term, ${\vec F}_j^{(1)}$,
 is the direct Coulomb repulsion stemming from neighboring macroions
and the plates
\begin{eqnarray}
{\vec F}_j^{(1)} = -{\vec \nabla}_{\vec R_j} \left( \sum_{i=1;
    j\not=i}^{N_m}  V_{mm} 
( \mid {\vec R}_i - {\vec R}_j \mid ) +
  V_{pm} (Z_j)\right)
\label{neu2}
\end{eqnarray} 
The second part ${\vec F}_j^{(2)}$ involves 
the electric part of the counterion-macroion interaction
and has the statistical definition
\begin{equation} 
{\vec F}_j^{(2)}=< \sum_{i=1}^{N_c} {\vec \nabla}_{\vec R_j}
  {{Zqe^2} \over {\epsilon  \mid {\vec R}_j - {\vec r}_i \mid }} >_c  
\label{8}
\end{equation} 
Finally, the third term ${\vec F}_j^{(3)}$ describes a
 depletion (or contact)  force arising from the hard-sphere part in
$V_{mc}(r)$, which can be expressed as an integral over the surface 
${\cal S}_j$ of the $j$th macroion 
\begin{equation}  
{\vec F}_j^{(3)}=k_BT \int_{{\cal S}_j} d{\vec f} \ \
\rho_c^{(0)}({\vec r}) 
\label{9}
\end{equation} 
where ${\vec f}$ is a surface  vector pointing towards the macroion center.
This depletion term is usually neglected in any DLVO or Poisson-Boltzmann treatment but
becomes actually important for strong macroion-counterion coupling. We
define the strength of Coulomb coupling via the dimensionless coupling 
parameter \cite{allah} 
\begin{equation}
\Gamma_{mc} ={Z \over q}{2 \lambda_B \over
  {d_m+d_c}} ,
\label{gamma_mc}
\end{equation}
 where the Bjerrum length is $\lambda_B = q^2e^2/\epsilon k_B T$.

A further interesting quantity is the
 counterion-averaged total potential energy defined as 
\begin{equation}
U( \{ {\vec R}_j \} ) = \sum_{i,j;i<j}^{N_m} V_{mm}(\mid
{\vec R}_i - {\vec R}_j \mid ) + < V_c >_c 
\label{neu5}
\end{equation} 
In general the effective force (\ref{neu4}) is different from the gradient of 
$U( \{ {\vec R}_j \} )$ \cite{Hartmut} i.e. 
\begin{equation}
{\vec F}_j \not= {\vec {\bar F}}_j \equiv -{\vec \nabla}_{\vec R_j}U( \{
{\vec R}_i \} ) 
\label{neu566}
\end{equation} 
In fact, as we shall show below these two quantities behave qualitatively different 
for strong coupling.  We emphasize that it is the effective force (\ref{neu4}) that is probed
in experiments.

\section{details of the computer simulation}

The Coulomb interactions involved in the primitive model 
are long-ranged but the periodically repeated system is finite
which poses a computational problem.
This can be solved in different ways. The simplest way to solve the
problem is to cut off the range of the Coulomb interaction by half of the
system size which is the minimum image convention (MIC). The MIC is
easy to implement but serious cut-off errors can be introduced.
A better way is to include $\cal N$ periodically repeated images (PRI) of neighbour
cells in $x$ and $y$ direction. Also the limit ${\cal N}\to\infty$ can be treated by
a suitable generalization of the traditional Ewald summation technique \cite{Ewald,han_pol,val_koh}
to a two-dimensional system. A straightforward
generalization, however, leads to quite
massive computational effort \cite{halley}. A much more effective 
alternative is the so-called Lekner summation method
 \cite{Lekner1,Lekner2} which has 
recently been applied successfully to the problem of effective
interactions between rodlike polyelectrolytes and like-charged planar
surfaces \cite{mashl12}.

A completely different way out of the problem is to study the system
on a surface of a four-dimensional (4D) hypersphere which itself is a 
compact closed geometry with spherical boundary conditions
\cite{izenberg}. Then one has to express the Coulomb forces in terms of the
appropriate 4D spherical coordinates which can be done
analytically, see Appendix A. Such spherical boundary conditions were
 effectively utilized in computer simulations of
 two-dimensional (2D) classical electrons \cite{han_lev,cal_lev} and other 2D 
 fluids \cite{caillol,kratky,cail_ela}. 
Simulations of the 3D system located on the surface of a 4D
 hypersphere were carried out for Lennard-Jones \cite{schrei}, 
hard sphere \cite{toboch} and charged \cite{cail1} systems. 
The hypersphere geometry (HSG) was also tested against Ewald
 summations to investigate the stability of charged
 interfaces \cite{cail2} and good agreement was found, 
even for strongly coupled interfaces. Simulations in HSG are 
much faster than that for Lekner sums or PRI as there is no sum over images.

In most of our investigations we have used  HSG simulations but tested them against
MIC, PRI and Lekner summations. Good agreement was found except for the MIC which
suffers from the early truncation of the Coulomb tail. We have performed Molecular Dynamics (MD)
simulations at room
temperature $T=293^o K$. A more
detailed description of the MD  procedure in HSG is given 
in Appendix B. The width of planar slit is fixed to be $2L=5 d_m$. 
 Different sets of system parameters  are summarized
in Table I.

\twocolumn[\hsize\textwidth\columnwidth\hsize\csname
@twocolumnfalse\endcsname

\begin{table}
\caption{Set of parameters used in our calculations}
\begin{tabular}{lccccccccc}
Run& $N_{m}$ & $Z$ & $q$ & $\sigma_{p}(e/cm^{2})$&
$\epsilon$&$\rho_m (1/cm^{3})$&$ d_m (cm)$&$ d_c (cm)$&$\Gamma_{mc}$ \\
\tableline
  A &   0  &  -   &  2   &  $ 1.24\times10^{11}$&
  78.3&$1.17\times10^{13}$& - & $5.32\times10^{-8}$&-  \\ 
  B &   1  &  200   &  2   &  $ 0.62\times10^{11}$&
  78.3&$1.17\times10^{13}$&$5.32\times10^{-6}$ & $5.32\times10^{-8}$&11  \\ 
  C &   1  &  200   &  2   &  $ 1.24\times10^{11}$&
  78.3&$varied$&$5.32\times10^{-6}$ & $5.32\times10^{-8}$&11  \\ 
  D &   1  &  100   &  2   &  $ 1.49\times10^{11}$&
  $varied $&$1.17\times10^{13}$&$5.32\times10^{-6}$ & $5.32\times10^{-8}$&$varied$ \\ 
  E &   1  &  100   &  2   &  $ 2.98\times10^{11}$&
  3.9&$1.17\times10^{13}$&$5.32\times10^{-6}$ & $5.32\times10^{-8}$&110  \\ 
  G &   1  &  100   &  2   &  $varied$&
  78.3&$9.36\times10^{16}$&$2.66\times10^{-7}$ & $2.66\times10^{-8}$&100  \\ 
  K &   1  &  32   &  2   &  $ 1.56\times10^{14}$&
  77.3&$1.9\times10^{18}$&$2.56\times10^{-7}$ & $2.56\times10^{-9}$&37  \\ 
  L &   2  &  200   &  2   &  $ 1.24\times10^{11}$&
  78.3&$2.34\times10^{13}$&$5.32\times10^{-6}$ & $5.32\times10^{-8}$&11  \\ 
  M &   2  &  100   &  2   &  $ varied$&
  3.9&$2.34\times10^{13}$&$5.32\times10^{-6}$ & $5.32\times10^{-8}$&110  \\ 
  N &   2  &  100   &  2   &  $varied$&
  78.3&$1.87\times10^{17}$&$2.66\times10^{-7}$ & $2.66\times10^{-8}$&100  \\ 
\end{tabular} 
\end{table}
]
 We take divalent counterions throughout our investigations.
The dielectric constant is that for water at room temperature $(\epsilon=78.3$) but
we have also  investigated cases where $\epsilon$ is smaller in order
to enhance the Coulomb coupling formally. The charge asymmetry $Z/q$ ranges from 
16 to 100. The
time step  $\triangle{t}$ of the simulation was typically chosen to be
$10^{-3}\,\sqrt{m\,{d^{3}_{m}}/e^{2}}$, (with $m$ denoting the mass of the counterions)
such that the reflection of
counterions following the  collision with the surface of macroions and
walls is calculated with high  precision. For every
run the equilibrium state of the system was checked during the
simulation time. This was done by monitoring the temperature, average velocity and
the distribution function of velocities and total potential energy of
the system. On average it took about $10^4$ MD steps to get into
equilibrium. Then during $5\cdot 10^4-5\cdot 10^5$ time steps, we gathered
statistics to perform the canonical averages for calculated quantities.

\section{counterion density profiles between charged plates}
\begin{figure}
   \epsfxsize=6cm
   \epsfysize=6cm 
   ~\hfill\epsfbox{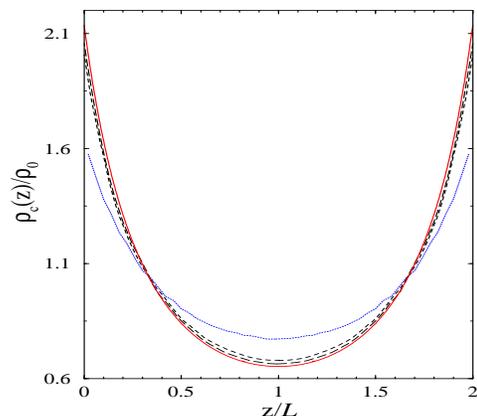}\hfill~
   \caption{Reduced counterion density profiles $\rho^{(0)}_c(z)/\rho_0$ versus 
reduced distance $z/L$. solid line- PB prediction and
  simulation result with incorporating Lekner summation method. Both data
do coincide on the same curve, long-dashed line-
  result of simulation in HSG, dashed line- result of PRI simulation
  with ${\cal N}=2$, dotted line- result of MIC simulation.}
   \label{fig1allah}
\end{figure}

First, as a reference case, let us  discuss the situation
without any macroion. This set-up   is well-studied
in the literature \cite{Robbins,Valleau}.
We consider equally charged surfaces  $\sigma_1 = \sigma_2 =\sigma_p$. 
 The imbalance in the
interaction with neighbours will push the  counterions toward the
plates. Consequently, a great
majority of neutralizing counterions reside within a thin surface
layer. For strong coupling, the width of this layer can be 
approximately expressed as \cite{rouzina}
\begin{equation}
\label{lambda}
\lambda_z = \frac {\lambda_{D}^{2}} {2L}, 
\end{equation}
where $\lambda_D$  is the bulk Debye screening length
\begin{equation}
\label{lambda2}
\lambda_{D}^{2}
= \frac {\epsilon k_B T} {4 \pi \rho_0 q^2 e^2} 
\end{equation}   
where $\rho_0\equiv \rho_c$.
Due to symmetry, the equilibrium 
counterion density profile only depends on $z$.
The analytical solution of the Poisson-Boltzmann (PB) equation for this profile is \cite{eng}
\begin{equation}
\label{hb}
\rho^{(PB)}_{c}(z)= \rho_c \frac {2\gamma_{0}^{2} \lambda_{D}^{2} }{L^{2}
  \cos ^2 (\gamma_{0} (1-{z \over L}))} 
\end{equation}
where $\gamma_0$ is defined via the solution of the implicit equation
\begin{equation}
\label{defgamma}
\frac {(L/\lambda_D)^{2}} {2\gamma_0} - \tan{\gamma_0} = 0 
\end{equation}
For parameters of moderate Coulomb coupling (run A), the PB result is shown
 as a solid line in Fig.\ref{fig1allah}.
 
The corresponding MD simulation data were obtained  with 600
counterions in the cell  using various boundary conditions. 
As expected, the PB theory coincides with the Lekner summation method
which treats best the long-range nature of the Coulomb interactions.
In  HSG the counterionic profiles are also very similar to the Lekner summation
while the MIC deviates significantly. The MIC can already be improved significantly
if ${\cal N}=2$ periodic repeated images are included. 
In conclusion, the agreement between Lekner summation
and HSG justifies the HSG a posteriori and gives evidence that the HSG produces
reliable results also for stronger couplings.

\section{Single macroion between charged plates}
\begin{figure}
   \epsfxsize=6.5cm
   \epsfysize=5cm
   ~\hfill\epsfbox{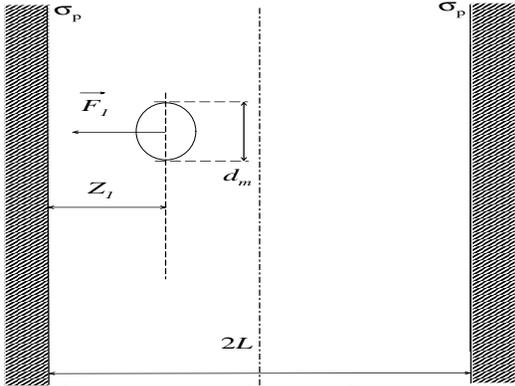}\hfill~
    \caption{Schematic picture for a single macroion between likely
  charged  planes of charge density $\sigma_p$, separated by distance
   $2L$.}
\label{fig2allah}
\end{figure}

Let us now consider a single macroion in the inter-lamellar
area. We put the macroion on the $z$-axis, such that its
position is at ${\vec R}_{1}=(0,0,Z_1)$. 
 A corresponding schematic picture is given in   Fig.\ref{fig2allah}.

The total force acting on the macroion only depends on $Z_1$ and points
along the $z$-axis.
Obviously, for the case $\sigma_1=\sigma_2=\sigma_p$ of symmetric plates considered here, the direct
part of the total force, ${\vec F}_j^{(1)}$, vanishes.
For the second (electrostatic) part, simple PB-theory applied to the case
of small macroion charge and small macroion diameter yields the following analytical expression
for the effective macroion force \cite{eng,anne}
\begin{equation}
\label{pbforce}
{\vec F}_{1}^{PB}=\frac {2 Z k_{B} T \gamma_{0}} {q L} \tan \left(\gamma_{0}
  \left(1-z/L \right) \right) {\vec e}_z
\end{equation}
where ${\vec e}_z$ is the unit vector in $z$-direction and $\gamma_0$
is given by (\ref{defgamma}). Note that only the counterion
density stemming from the charged plates
has to be inserted in (\ref{defgamma})
i.e. $\rho_o=\frac{\sigma_p}{L\mid qe \mid}$.
This force pushes the macroion
 towards the mid-plane, i.e.\ the wall-particle interaction is repulsive.

\begin{figure}
   \epsfxsize=6cm
   \epsfysize=7cm 
   ~\hfill\epsfbox{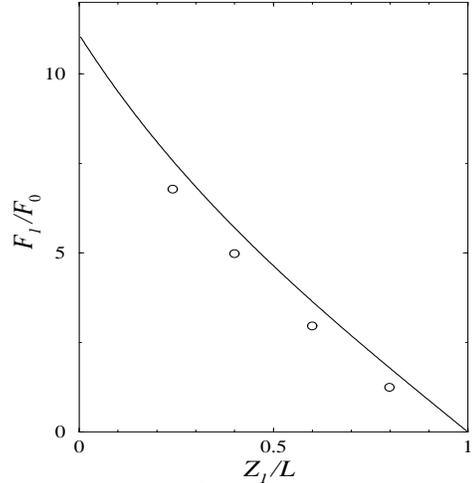}\hfill~
\caption{Force $F_1={\vec F}_1 \cdot {\vec e}_z$ acting  on a single macro-ion versus
 reduced  macro-ion distance $Z_1/L$. The force is scaled by the (arbitrary)
unit $F_{0}= \frac{Zqe^{2}} {\epsilon {d_{m}^{2}}}$. 
 The system parameters are from run B\@. The solid line is the
prediction from PB theory. The open circles are  simulation results in HSG\@. The statistical error corresponds
to the symbol size.}
\label{fig3allah}
\end{figure}

\begin{figure}
   \epsfxsize=6cm
   \epsfysize=7cm 
   ~\hfill\epsfbox{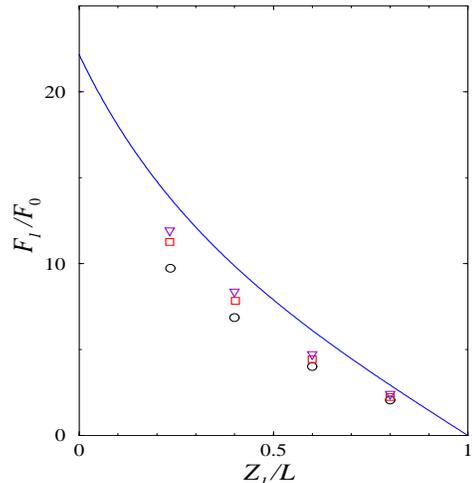}\hfill~
\caption{Same as Fig.\ref{fig3allah} but now for run C.  Symbols are
  simulation results in HSG for various macroion number densities:
 circles: $\rho_m = 1.17\times 10^{13}  cm^{-3}$, squares: $\rho_m =
  2.0\times 10^{12} cm^{-3}$, triangles: $\rho_m = 1.0\times 10^{12} cm^{-3}$.}
\label{fig4allah}
\end{figure}

\begin{figure}
   \epsfxsize=6cm
   \epsfysize=7cm 
   ~\hfill\epsfbox{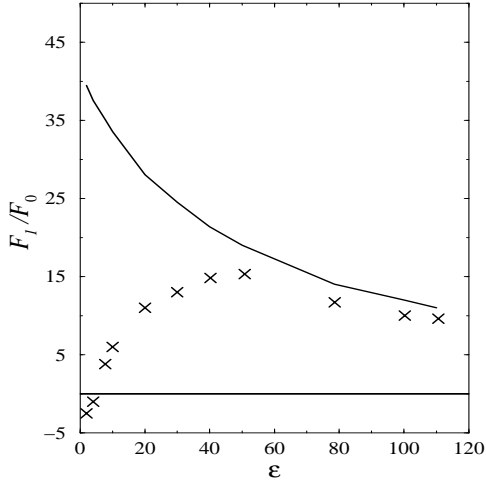}\hfill~
\caption{Force $F_1={\vec F}_1 \cdot {\vec e}_z$ acting  on a single macro-ion versus
dielectric constant $\epsilon$ for a fixed position at $Z_1 = d_m $.
The force is scaled by the (arbitrary)
unit $F_{0}=0.01 \frac{Zqe^{2}} {\epsilon {d_{m}^{2}}}$. The parameters of the
  system are from run D. The solid line is the  prediction of PB
  theory. The crosses are simulation results in HSG.}
\label{fig5allah}
\end{figure}

The expression (\ref{pbforce}) will break down, however, for a large
macroion diameter $d_m$ and for strong 
macroion-counterion coupling parameter $\Gamma_{mc}$.
We have tested the PB-theory against ``exact" computer simulation data.
 For moderate couplings (run B and run C) the results for the total force
 $F_1={\vec F}_1 \cdot {\vec e}_z$ are shown in
 Figs.\ref{fig3allah}-\ref{fig4allah}. In Fig.\ref{fig3allah}, a surprising agreement between theory and simulation
is obtained despite the fact that the macroion charge is large.
 This justifies the theoretical
conclusions drawn in Refs.\ \cite{anne,Jay} based on PB theory. The deviation between theory and 
simulation are larger in Fig.\ref{fig4allah} where the surface charge density was doubled.
Here, also the system size dependence (resp.\ the dependence on the
macroion density) was studied in the simulation.
  As expected the agreement becomes better for
a larger system size (resp.\ for a smaller macroion density) since the theory is constructed formally
for vanishing macroion density. In addition, we repeated the
calculations for $\rho_m = 1.17\times 10^{13} cm^{-3}$ (corresp.\ to
the circles in Fig.\ref{fig4allah}) using the PRI method with $N=4$
and got the same results as in HSG\@.
 We now enhance the Coulomb coupling by formally reducing the dielectric constant
$\epsilon$. For a fixed macroion position at $Z_1=d_m$, the force $F_1$ is shown in Fig.\ref{fig5allah}
for the parameters of run D.
 While the PB theory always predicts a repulsive force, 
the simulation data
are in accordance with theory only for large $\epsilon$ but the force changes its sign 
for $\epsilon <10$. Hence as expected the theory breaks down for large Coulomb coupling
where correlation between the counterions become significant.

\begin{figure}
   \epsfxsize=6cm
   \epsfysize=7cm 
   ~\hfill\epsfbox{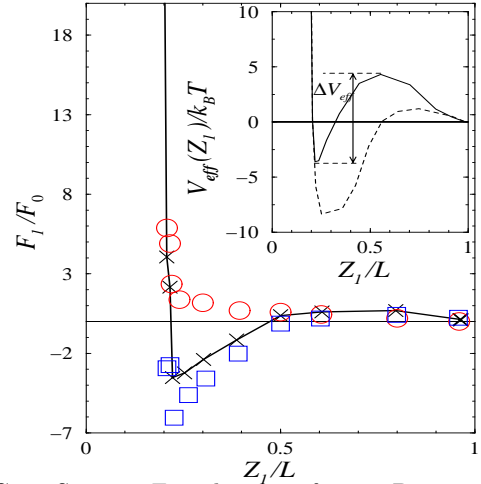}\hfill~
\caption{Same as Fig.\ref{fig3allah} but now for run D, $\epsilon=3.9$, and for a force unit
of $F_{0}=0.1 \frac{Zqe^{2}} {\epsilon {d_{m}^{2}}}$. The crosses are 
simulation data in HSG for the total force $F_1$, the  squares (resp.\ circles)
are simulation data the electrostatic part  $F_1^{(2)}$ (resp.\ the depletion
part $F_1^{(3)}$).The line is a guide
to the eye for the total force. The inset shows the effective potential in units of
$k_BT$ versus
  reduced  macro-ion distance $Z_1/L$ together with the energy barrier
  $\Delta V_{eff}$. The solid line is for run D with
$\epsilon =3.9$, dashed line is for run E.}
\label{fig6allah}
\end{figure}

 For the same run D, the distance-resolved  macroion force $F_1$ is shown in Fig.\ref{fig6allah} for
 a strongly reduced dielectric constant $\epsilon=3.9$.
 The simulation data
were obtained in HSG but confirmed by PRI calculations with ${\cal N}=4$.
 The electrostatic part 
$F_1^{(2)}={\vec F}_1^{(2)} \cdot {\vec e}_z$ and the depletion 
part $F_1^{(3)}={\vec F}_1^{(3)} \cdot {\vec e}_z$ are shown separately. $F_z^{(3)}$ is
always  repulsive and increases with decreasing $Z_1$, at least if the macroion is not
too close to the surface when the counterion depletion between the macroion and the wall
induced by the finite  counterion core is negligible. This
 is an  expected behavior, since in general there are more counterions close to the walls.
 The pure electrostatic contribution, $F_1^{(2)}$, on the other hand,
 exhibits  a more subtle behavior. If the macroion is close to the midplane, 
it is repulsive, then it becomes attractive as the macroion is getting closer to the plates.
 As a function of macroion distance,
the total force $F_1$ is repulsive, attractive and becomes repulsive again.
For  small separations (which are still larger than the microscopic counterionic core)
the force is dominated by the repulsive depletion force. Hence the
 macroion has got three equilibrium positions, two of them are stable, namely 
the midplane  and  a position in the vicinity of the plate. In order to extract 
more information, we have calculated the effective wall-particle potential
defined by
\begin{equation}
V_{eff}(Z_1) = - \int_{0}^{Z_1} F_1(h) dh 
\label{eff}
\end{equation}
by integrating our data with respect to the macroion altitude $h$. This
quantity  is shown as an inset in Fig.\ref{fig6allah}.
One first sees that the global minimum is in the vicinity of the walls. Furthermore
the barrier height $\Delta V_{eff}$ to escape from there is about  $8 k_BT$. This
implies that the time for a colloidal particle to escape
from the position close to the surface is roughly $\tau_0 \exp \left(\Delta
V_{eff}/k_BT \right) = e^8 \tau_0\approx 3000 \tau_0$ \cite{Russel,Sauer}
where $\tau_0$ is a Brownian time scale governing the decay of
dynamical correlations of the macroion. It can also be seen that, for a
doubled surface charge (run E), the height of barrier increases.

\begin{figure}
   \epsfxsize=6cm
   \epsfysize=7cm 
   ~\hfill\epsfbox{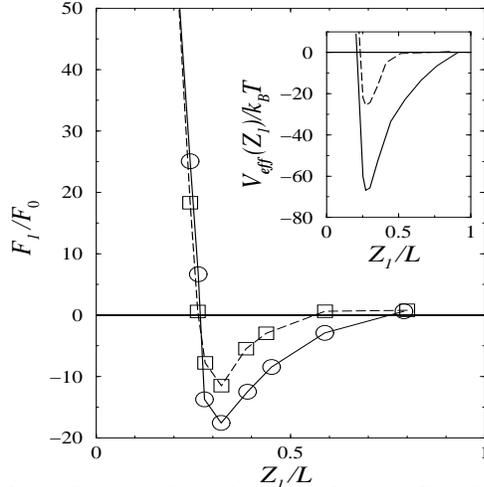}\hfill~
\caption{Same as Fig.\ref{fig6allah} but now for run G and for a force unit
of $F_{0}=0.1 \frac{Zqe^{2}} {\epsilon {d_{m}^{2}}}$. The squares are
simulation results for the total force in HSG
for $\sigma_p
  =1.19\times10^{14} { e \over cm^{2}} $, while the circles are for 
$\sigma_p  =2.38\times10^{14} {e \over cm^{2}}$. The lines are a guide
to the eye.  The inset shows the effective potential in units of
$k_BT$ versus
  reduced  macro-ion distance $Z_1/L$. The dashed line is for $\sigma_p
  =1.19\times10^{14} { e \over cm^{2}} $, the solid line is for $\sigma_p  =2.38\times10^{14} {e \over cm^{2}}$.}
\label{fig7allah}
\end{figure}

A similar behaviour occurs for another parameter combinations (run G), see
Fig.\ref{fig7allah}, corresponding to aqueous suspensions of micelle-sized macroions. 
It hence seems to be a generic feature of the primitive model for
strong Coulomb coupling.
\begin{figure}
   \epsfxsize=6cm
   \epsfysize=7cm 
   ~\hfill\epsfbox{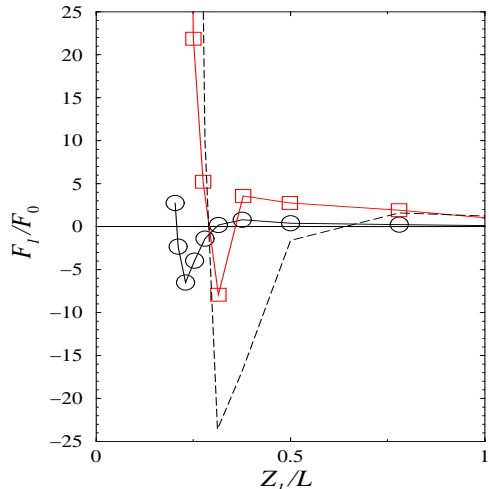}\hfill~
\caption{Effective force $F_1={\vec F}_1 \cdot {\vec e}_z$ (circles) and gradient
of the potential energy $\bar F_1={\vec {\bar F}}_1 \cdot {\vec e}_z$ (squares) versus
 reduced  macro-ion distance $Z_1/L$ for run K. The unit of the force
 is  $F_{0}=0.1 {{Zqe^{2}} \over {\epsilon d_{m}^{2}} }$. the lines
 are a guide to the eye. The dashed line are data from Ref.[60]}    
\label{fig8allah} 
\end{figure}
 We note that the barrier height $\Delta V_{eff}$ is about $70
k_B T$ which implies a very large escape time. Finally we show for a certain parameter combination (run K) which
was also used in Ref.\  \cite{bo} that the averaged force ${\vec F}_1$ differs from the
gradient of the averaged potential energy ${\vec {\bar F}}_1$. As in 
Ref.\  \cite{bo}, the system consists of
a single macroion in a planar slit of width $5d_m$, with one charged and one neutral 
wall. Results are given in Fig.\ref{fig8allah}.
 We conclude that  the forces
behave  even qualitatively different. The average force  ${\vec F}_1$
is a short-range attractive force which becomes repulsive only for
touching macroion configurations. Contrary to that, the force  ${\vec {\bar
F}}_1$ is repulsive up to distance about $d_m/2$ from the wall
surface. Note that our data actually differ from those of Ref.\
\cite{bo} due to the early truncation of the Coulomb interaction performed
there.    

\section{Two macro-ions between plates}
\begin{figure}
   \epsfxsize=7.5cm
   \epsfysize=6cm 
   ~\hfill\epsfbox{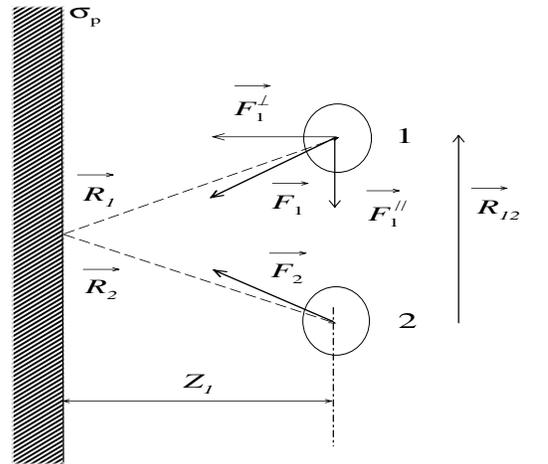}\hfill~
\caption{Schematic  picture for the macroion  pair near a charged wall of
 surface charge density
   $\sigma_p$. For the sake of clarity, the position of second wall
   is omitted. The different forces are shown for the case of a mutual attraction.}
\label{fig9allah}
\end{figure}

We finally  consider two equally charged macroions at the positions
${\vec R}_{1}=(X_1 ,Y_1 ,Z_1)$ and ${\vec R}_{2}=(X_2 ,Y_2 ,Z_2)$ confined
between plates. A schematic picture is given in Fig.\ref{fig9allah}.

 In order
to reduce the parameter space, we assume for
simplicity that both  macroions have the same  altitude $Z_1=Z_2$. The
distance between the macroion centers is $R_{12} =\mid {\vec R}_1 -
{\vec R}_2 \mid $ where the difference vector  ${\vec R}_{12}={\vec R}_1 -
{\vec R}_2$ is in the $xy$-plane.  We assume that only one of the walls is charged and that
the second wall is neutral. This gives us the
possibility to simulate higher surface charge densities. Also for strong coupling,
the counterions of the two different walls are practically decoupled such that
the set-up with a single charged wall is not expected to differ much from the symmetrical set-up.
The total force acting on the two macroions can be split into a part 
pointing in $z$-direction and another contribution pointing along ${\vec R}_{12}$.
Hence we write ${\vec F}_j = {\vec F}^\|_j + {\vec F}^\bot_j$ defining
${\vec F}^\|_j= \left({\vec F}_j \cdot {\vec R}_{12} \right) \cdot
{\vec R}_{12}/R_{12}^2$ and 
${\vec F}^\bot_j= \left( {\vec F}_j \cdot {\vec e}_z \right) \cdot {\vec e}_z$ for $j=1,2$. Clearly,
${\vec F}^\bot_1={\vec F}^\bot_2$, and ${\vec F}^\|_1=-{\vec F}^\|_2$.

It is instructive to compare these force to the DLVO bulk theory which yields
\begin{eqnarray}
{\vec F}^{DLVO}_1 =&& \frac {{Z}^2 e^2 \exp{((d_m - R_{12})/ \lambda_{D}) }
  } {\epsilon R_{12}(1+ {d_m/2 \lambda_D})^2} \nonumber\\
&&\times \left({1 \over R_{12}}+{1
  \over \lambda_D}
  \right) {{\vec R}_{12} \over R_{12}}
\label{dlvo}
\end{eqnarray}
Here the Debye screening length $\lambda_D$ is given by eq.(\ref{lambda2}), where $\rho_0$
corresponds to the counterion number density coming only from the
macroions, $\rho_{0} = {Z \over {q}}\rho_{m}$. One can also modify
the DLVO theory by admitting screening also from the counterions stemming from the wall
assuming they follow a Poisson-Boltzmann density profile.
This yields the PB force \cite{anne,Jay}
\begin{equation}
{\vec F}^{PB}_1 = ({\vec F}^{PB}_1)^\| + ({\vec F}^{PB}_1)^\bot 
\label{PBA}
\end{equation}
where we get for the parallel part of the force
\begin{eqnarray}
\label{PBpar}
({\vec F}_1^{PB})^\| =&& \frac {{Z}^2 e^2 \exp{(- R_{12}/ \lambda_{D}(Z_1))}
  } {\epsilon R_{12}} \nonumber \\
&&\times \left({1 \over R_{12}}+{1
  \over \lambda_D(Z_1)}\right) {{\vec R}_{12} \over R_{12}}.
\end{eqnarray}
The perpendicular part of the force is
\begin{eqnarray}
\label{PBper}
({\vec F}^{PB}_1)^\bot =&&  {\vec F}_1^{PB} - \frac {{Z}^2 e^2 }
   {\epsilon} \frac{\lambda_D(Z_1) \gamma_0^3}{L^3} \nonumber \\
&& \times \frac {\tan
   \left(\gamma_0 (1-Z_1/L)\right)} {\cos^2
   \left(\gamma_0 (1-Z_1/L)\right)}  {\vec e}_z  
\end{eqnarray}
\begin{figure}
   \epsfxsize=6cm
   \epsfysize=7cm 
   ~\hfill\epsfbox{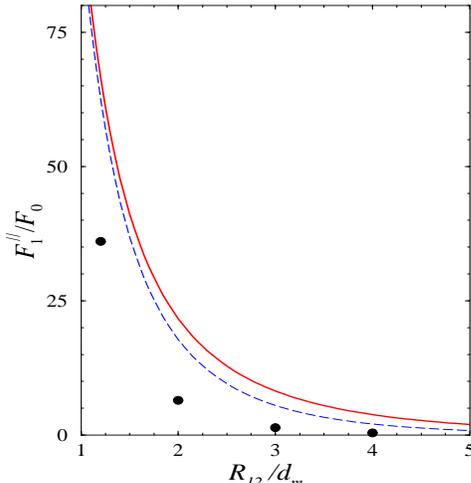}\hfill~
\caption{Parallel part of the effective force acting onto a macroion pair,
 $F^\|_1={\vec F}^\|_1\cdot {{\vec R}_{12} / R_{12}}$, versus
reduced interparticle distance $R_{12}/d_m$. The unit of the force is $F_{0}=\left( 
\frac{Z^{2} e^{2}} {\epsilon {d_{m}^{2}}}\right )\,\times 10^{-2}$. The
  parameters of system are from run L and for an altitude of macroions of
  $Z_1=0.6 d_m$. The solid line is the bulk DLVO theory, the
dashed line is the Poisson-Boltzmann theory (\ref{PBpar}) and the points are  simulation results
in HSG\@. The statistical error corresponds
to the symbol size.}
\label{fig10allah}
\end{figure}
The space dependent Debye screening length is 
\begin{equation}
\label{ldz}
\lambda_D(Z_1) = \left( 4 \pi \lambda_B \left( {Z \over q}\rho_m +
\rho_c^{PB}(Z_1)\right)     \right)^{-\frac{1}{2}}
\end{equation}
Here $\rho_c^{PB}(z)$ and ${\vec F}_1^{PB}$ are given by Eqns.\ (\ref{hb}) and (\ref{pbforce}).  
Contrary to the bulk DLVO force, the  PB force 
has an additional perpendicular part for a pair of particles (second
term in (\ref{PBper})). This additional force is
attractive. Still the total perpendicular force
(\ref{PBper}) is always repulsive.

  For the parameters of run L corresponding to weak coupling,  simulation results 
for $F^\|_1={\vec F}^\|_1 \cdot {{\vec R}_{12} / R_{12}}$ are shown in
Fig.\ref{fig10allah}.

 The solid line corresponds to the bulk  DLVO force, and
the dashed line is the Poisson-Boltzmann result.
The force is repulsive both in theory and simulation,
but the theories overestimate the force significantly. As expected
the Poisson-Boltzmann approach yields better agreement than DLVO bulk theory.

\begin{figure}
   \epsfxsize=6cm
   \epsfysize=7cm 
   ~\hfill\epsfbox{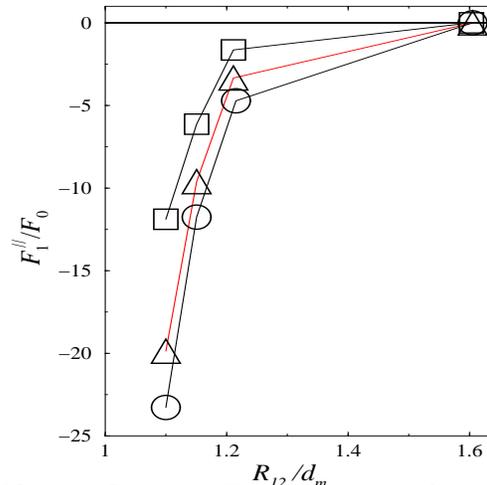}\hfill~
\caption{Same as Fig.\ref{fig10allah} but now for run M and $F_{0}=\left( 
\frac{Z^{2} e^{2}} {\epsilon {d_{m}^{2}}}\right )\,\times 10^{-3}$.
 Simulation results are shown for three different surface charges:
 squares: $\sigma_p = 0 {e \over cm^{2}}$, triangles: $\sigma_p =
 2.98\times10^{11} {e \over cm^{2}}$, circles: $\sigma_p = 5.95\times10^{11}
 {e \over cm^{2}}$. The lines are a guide to the eye.}
\label{fig11allah}
\end{figure}

\begin{figure}
   \epsfxsize=6cm
   \epsfysize=7cm 
   ~\hfill\epsfbox{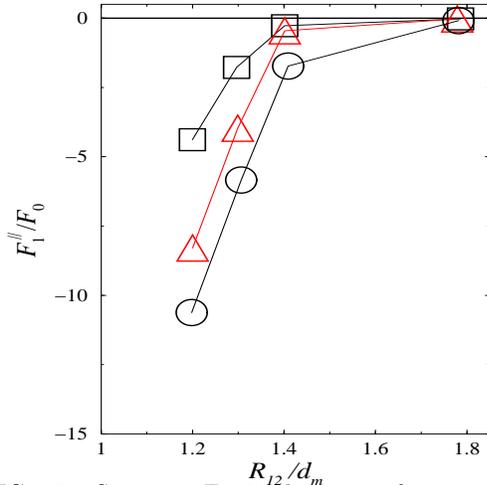}\hfill~
\caption{Same as Fig.\ref{fig11allah} but now for run N and for 
  $Z_1 =0.7\, d_m$. Results are shown for three different surface charges:
 squares: $\sigma_p = 0 {e \over cm^{2}}$, triangles: $\sigma_p =
 1.19\times10^{14} {e \over cm^{2}}$, circles: $\sigma_p = 2.38\times10^{14}
 {e \over cm^{2}}$.}
\label{fig12allah}
\end{figure}

Results for $F^\|_1$ for stronger coupling (runs M and N) are
displayed in Figs.\ref{fig11allah}-\ref{fig12allah}.
 For a neutral wall, the
interaction force between macroions is already attractive.
 With increasing  surface charge the attraction
between macroions becomes stronger. Clearly this attraction is neither contained
in DLVO theory nor in the Poisson-Boltzmann approach (\ref{PBpar}).

\begin{figure}
   \epsfxsize=6cm
   \epsfysize=7cm 
   ~\hfill\epsfbox{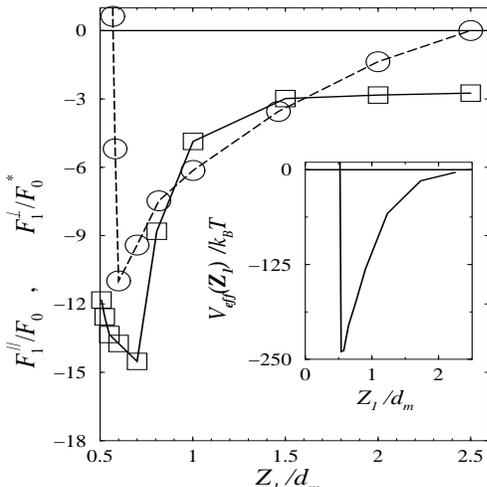}\hfill~
\caption{Parallel $F^\|_1={\vec F}^\|_1\cdot {{\vec R}_{12} /
 R_{12}}$ ( squares) and perpendicular
 $F^\bot_1={\vec F}^\bot_1\cdot {\vec e}_z$ (circles) parts of effective force versus reduced altitude $Z_1/d_m$
for fixed interparticle spacing $R_{12}=1.2\,d_m$. The unit of the
 force $F^\|_1$ is $F_{0}=\left( 
\frac{Z^{2} e^{2}} {\epsilon {d_{m}^{2}}}\right )\,\times 10^{-3}$,
 and for the force  $F^\bot_1$  is $F_{0}^*=\left( 
\frac{Z^{2} e^{2}} {\epsilon {d_{m}^{2}}}\right )\,\times 10^{-2}$.  The surface charge
 density is $\sigma_p = 2.38\times10^{14} {e \over cm^{2}}$. The inset shows the effective potential in units of
$k_BT$ versus
  reduced  macro-ion distance $Z_1/d_m$. }
\label{fig13allah}
\end{figure}

In Fig.\ref{fig13allah} we fixed the macroion distance and calculated $F^\|_1$
and the force perpendicular to the plates,
$F^\bot_1={\vec F}^\bot_1 \cdot {\vec e}_z$ versus altitude $Z_1$ for
run N.

\begin{figure}
   \epsfxsize=6cm
   \epsfysize=7cm 
   ~\hfill\epsfbox{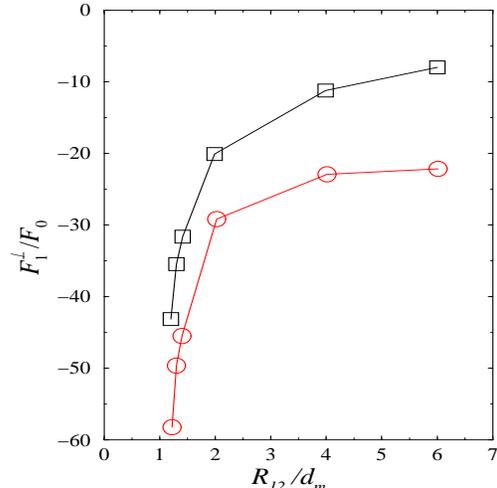}\hfill~
\caption{Perpendicular part of the effective force acting onto a macroion pair,
$F^\bot_1={\vec F}^\bot_1\cdot {\vec e}_z$ in units of $F_{0}=\left( 
\frac{Z^{2} e^{2}} {\epsilon {d_{m}^{2}}}\right )\,\times 10^{-3}$ versus
 dimensionless interparticle  distance  $R_{12}/d_m$. The
  parameters of system are from run N and the altitude of macroions is
fixed to $Z_1=0.7\, d_m$. Simulation results are shown for two different surface charges:
 squares: $\sigma_p =
 1.19\times10^{14} {e \over cm^{2}}$, circles: $\sigma_p = 2.38\times10^{14}
 {e \over cm^{2}}$.}
\label{fig14allah}
\end{figure}

 There is attraction. Both the interparticle attraction and the
wall-particle attraction become stronger
in the vicinity of the plate. The effective wall-particle
interaction potential for the perpendicular part
is shown as an inset in Fig.\ref{fig13allah}. Note
that the minimum of $V_{eff}$ is much more than twice as large as in
the single macroion case (compare to inset in
Fig.\ref{fig7allah}, solid line). Thus, a pair of macroions near a
planar surface is more stable than a single macroion. This is also
evident from the results for run N shown 
in Fig.\ref{fig14allah}.
 Again there is attraction towards the plate
for varied $R_{12}$ and fixed $Z_1$. 
The attraction
becomes stronger if the interparticle distance is decreasing. This shows that the attraction
between the wall and a single macroion discussed in chapter V is stable and even enhanced
if more macroions are close to the wall. This leads us to the final conclusion that
the macroions will assemble on top of the surface forming two-dimensional colloidal
layers.

\section{Conclusions}
We have simulated the effective force between macroions 
confined in a slit-geometry. An effective attraction was found for
strong Coulomb coupling. In particular, the effective potential of a single macroion confined
between two parallel charged plates was found to have two stable minima
where the total force vanishes: the first is in the mid-plane, the second close
to the walls. This result was confirmed for two macroions. In this
case the attraction
towards the walls was even stronger than for a single macroion.
Our most important conclusion is that the attractive force will result
in two-dimensional colloidal layers on top of the plates. As the depth
of the attractive potential is larger than $k_B T$, these layers
possess a large life-time with respect to thermal fluctuations. The layers
should be crystalline as the interparticle interaction is also attractive.
This can explain at least qualitatively the long-lived metastable
crystalline layers found in recent experiments on confined samples
of charge colloidal suspensions \cite{Larsen_crystallites,larsen}.

We want to add some remarks: First, our parameters are actually
different form those describing the experiments. The main difference is the
high surface charge of the glass plates within an area spanned by a typical macroion separation
distance. Such a system cannot be simulated since it
requires a huge number of counterions in the simulation box. We have mimicked the high
surface charge by dealing with a small dielectric constant, but strictly speaking
this corresponds to a different  system. Second, the mechanism of our attraction is similar
to that proposed recently by us in the bulk case \cite{allah}. It only occurs for strong coupling
with divalent counterions and is short-ranged. In this respect, it behaves different than in
experiment where the attraction was long-ranged. We emphazise again that the depletion force
is crucial in the strong coupling parameter regime. 
Third, our computer simulation data were tested
against simple DLVO- or Poisson-Boltzmann-type theories. 
It would be interesting to use them as benchmark data for more sophisticated theoretical 
approaches which predict attraction as e.g.\ the density functional perturbation theory
recently proposed by Goulding and Hansen \cite{dave23}.
Finally, in our simulations, we neglected any impurity or added salt ions. Their inclusion
increases substantially the number of microscopic particles and would
lead to more extensive simulation. A further challenge would be to incorporate image charges properly into the model which requires a non-trivial extension of our approach.

\acknowledgments
We  thank D. Goulding, J. P. Hansen, C. N.  Likos for helpful comments. Financial
support from the Deutsche Forschungsgemeinschaft within (SFB 237 and Schwerpunkt
Benetzung und Strukturbildung an Grenzfl{\"a}chen)
is gratefully acknowledged. We also thank the IFF at the Forschungszentrum
J{\"u}lich for providing CPU time.

\appendix
\section{Definition of forces in Hyper Sphere Geometry}
\begin{figure}
   \epsfxsize=6cm
   \epsfysize=7.5cm 
   ~\hfill\epsfbox{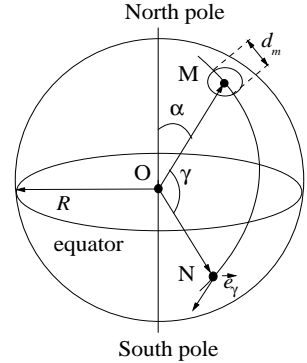}\hfill~
\caption{Schematic view of the hypersphere (projected to three dimensions)
illustrating the angular coordinate
  $\alpha$.}
\label{fig15allah}
\end{figure}

\begin{figure}
   \epsfxsize=6cm
   \epsfysize=7.5cm 
   ~\hfill\epsfbox{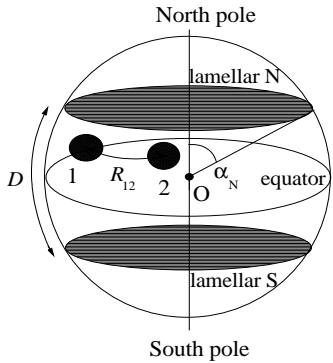}\hfill~
\caption{Schematic view of the hypersphere (projected to three dimensions) illustrating
a situation with two parallel charged
  interfaces.}
\label{fig16allah}
\end{figure}

We shortly present here some technical details in HSG within the
primitive model. For more details, we refer to Ref.\ \cite{cail2}.\\
The charged hard spheres are confined on the surface
$S_{3}$ of a 4D hypersphere. Without confinement, in the bulk,
the whole surface of the hypersphere is accessible to the particles. Since
 $S_{3}$ is compact, the total charge in a closed space must be equal
to zero. We define a pseudocharge as the association of a point charge
$q_i$ located at the point say $M$, and a neutralizing background of
total charge $-q_i$. The position of pseudocharge is specified by the 4D spherical coordinates
$(\alpha, \theta, \varphi)$ (see  Fig.~\ref{fig15allah}).
 Then the
Cartesian components of the unit vector ${\vec u}(M)=\vec{OM}$/$R$ reads
\begin{eqnarray}
&& u_{1}=\sin{\alpha}\sin{\theta}\cos{\varphi},\,\,
u_{2}=\sin{\alpha}\sin{\theta}\sin{\varphi},\,\,\nonumber \\
&& u_{3}=\sin{\alpha}\cos{\theta},\,\, u_{4}=\cos{\alpha}
\end{eqnarray}
Here $R$ is the hypersphere radius, $O$ is the center of the hypersphere
and  the angle $\alpha$ determines the
distance $R\alpha$ from north pole, see Fig.~\ref{fig15allah}. The distance
between two pseudocharges $q_{i}$ (at point $M$) and $q_{j}$ (at point
$N$) is measured along the geodesic line joining these points
\begin{equation}
r_{MN} = R\,\gamma ,
\end{equation}
where $\gamma$ is the angle between vectors $\vec{OM}$ and $\vec{ON}$,
\begin{equation}
\gamma=\arccos{\frac{\vec{OM}\,\cdot \vec{ON}}{R^{2}}}
\label{gamma}
\end{equation}
 The Coulomb force
$\vec{F}_{ij}$ between  pseudocharge $q_{i}$ and  pseudocharge $q_{j}$ is 
\begin{equation}
{\vec{F}}_{ij} = \frac{q_{j}\,q_{i}}{\epsilon \pi
 R^{2}}\left(\cot{\gamma}+\frac{\pi
 - \gamma}{\sin^{2}{\gamma}} \right ){\vec{e}}_{\gamma}(N).
\label{4dforce}
\end{equation}
Here $\vec{e}_{\gamma}(N)$ denotes the unit vector tangent to geodesic $MN$ at
 point $N$
\begin{equation}
\vec{e}_{\gamma} = -\frac{1}{\sin{\gamma}}{\vec u}(M) +
\cot{\gamma}\,\,{\vec u}(N)
\end{equation}
For short distances $r_{MN}$, there is the hard core repulsion.
 A hard sphere of diameter $d_{i}$ ($i=c,m$) centered
around the  point $M$ on $S_{3}$ is defined as the set of points $M_{0}$ such
that $R\,\gamma_{MM_{0}}= R\,\arccos{\frac{\vec{OM}\,\cdot
\vec{OM_0}}{R^{2}}} < d_{i}/2$. Thus, 
 the hard sphere potential between two pseudocharge is defined by 
\begin{equation}
U_{ij}=\left\{ \begin{array}{ll}
\infty & \text{if}\: \gamma \langle \frac{d_{i}+d_{j}}{2R},\,\, i,j=c,m \\
0 & \text{otherwise.} 
\end{array}
\right.
\end{equation}
Let us now consider the mixed hard sphere system to be confined
between two charged walls. On $S_{3}$ geometry it corresponds to the
two charged lamellae, parallel to each other, localized symmetrically
on opposite side of the equator (see Fig.(\ref{fig16allah})).
 They result
from a conical section of the hypersphere, with angular apertures
equal to $\alpha_{N}$ for north lamellar and $\alpha_{S}=\pi -
\alpha_{N}$ for south lamellar. The area of
each lamellar is\,\, $S_{p}=4\pi R^{2}\sin^{2}{\alpha_{N}}$ and
the volume confined between lamellae  is given by
\begin{equation}
V(\alpha_{N}) =  \pi R^{3}(2\alpha_{N} - \sin{2\alpha_{N}})
\end{equation}
Then the separation distance between lamellar is $D=R(\pi -
2\alpha_{N})$. For the symmetrical case considered in this paper,
when both lamellar are charged equally with surface charge density
$\sigma_{p}$, the charge electroneutrality of simulation cell together
with eq.~(\ref{neutral1}) requires 
\begin{equation}
N_{p}\, q_{c}\, + \,2\,Q_{p} = 0.
\end{equation}
Here $Q_{p}=\sigma_{p} S_{p}$ is the net charge of one lamellar,
 \,$N_{p}$ is the number of counter-ions coming from planes. 

Finally, we give the definition of interaction forces. 
The lamellar-lamellar  repulsion force is 
\begin{equation}
F_{pp} = \frac{Q_{p}^{2}}{\epsilon\,\pi\,R^{2}}\left(-\cot{\alpha_{N}}+\frac{\alpha_{N}}{\sin^{2}{\alpha_{N}}}\right).
\end{equation}
The ion-ion repulsion ($i=j$) and attraction ($i\neq j$) force is
\begin{equation}
F_{ij} = \frac{q_{j}\,q_{i}}{\epsilon\,\pi\,R^{2}}\left(\cot{\gamma}+\frac{\pi - \gamma}{\sin^{2}{\gamma}}\right),
\end{equation}
here $\gamma$ is given by (\ref{gamma}) and the particles $i$ and $j$ are at
points $M$ and $N$, $i,j=m,c$.  
The ion-lamellar repulsion ($i=m$) and attraction ($i=c$) force is

\begin{equation}
F_{ip} = \frac{Q_{p}\,q_{i}}{\epsilon\,\pi\,R^{2}}\left(\cot{\alpha_{i}}+\frac{\pi\,-
\alpha_{i}}{\sin^{2}{\alpha_{i}}}\right).
\end{equation}
for the upper lamella and 

\begin{equation}
F_{ip} =
\frac{Q_{p}\,q_{i}}{\epsilon\,\pi\,R^{2}}\left(\cot{\alpha_{i}}-\frac{\alpha_{i}}
{\sin^{2}{\alpha_{i}}}\right).
\end{equation}
for the second lamella.
The direction of the forces is always along the geodesic line.

The lamellar-ion hard core repulsion becomes
\begin{equation}
U_{ip}=\cases {\infty & for $ \alpha_{i} < \alpha_{N}+d_{i}/2R,\,\,i=c,m $ \cr
0 & otherwise\cr} 
\end{equation}


\section{equation of motion for single counter-ion in hyperspherical geometry}
In this section we translate  Newton equation of motion onto HSG for a
counterion in an external electrical field created by charged planes, other
counterions and fixed macroions. First of all, let us define the
displacement of counterion at
point $M$ on $S_{3}$. The differential $d\,\vec{OM}$ of vector
$\vec{OM}$ ($M$ remaining on the surface of the hypersphere) is 
\begin{equation}
d\,\vec{OM} =
R\,d\alpha\,\vec{e}_{\alpha}+R\,\sin{\alpha}\,d\theta\,\vec{e}_{\theta}+
R\,\sin{\alpha}\,\sin{\theta}\,\vec{e}_{\varphi}
\end{equation}
where the unit vectors
[$\vec{e}_{\alpha},\vec{e}_{\theta},\vec{e}_{\varphi}$]
given by
\begin{eqnarray*}
\vec{e_{\alpha}}&=&(\cos{\alpha}\sin{\theta}\cos{\varphi},\cos{\alpha}\sin{\theta}\sin{\varphi},
 \nonumber \\
&& \cos{\alpha}\cos{\theta},-\sin{\theta}),
\end{eqnarray*}

\begin{eqnarray}
\vec{e_{\theta}}=(\cos{\theta}\cos{\varphi},\cos{\theta}\sin{\varphi}
,-\sin{\theta},0),
\end{eqnarray}

\begin{eqnarray*}
\vec{e_{\varphi}}=(-\sin{\varphi},\cos{\varphi},0,0).
\end{eqnarray*}
 constitute an
orthogonal frame at point $M$. For the kinetic energy in  terms of the
variables ($\alpha,\theta,\varphi$) we get
\begin{eqnarray}
T= && \frac{mv^2}{2}=\frac{m}{2}\left( \frac{d\vec{OM}}{dt}
\right)^{2}  \nonumber \\ 
&& =\frac{m}{2} \left(R^{2}\,{\dot{\alpha}}^{2} +
  R^{2}\,{\sin{\alpha}}^{2}\,{\dot{\theta}}^{2}
+ R^{2}\,{\sin{\alpha}}^{2}\,{\sin{\theta}}^{2}\,{\dot{\varphi}}^{2} \right)
\label{kin}
\end{eqnarray}
For the potential energy we have relations
\begin{eqnarray}
-\frac{\partial{U}}{\partial{\alpha}} = R\,F_{\alpha}, \,\,\,&&
-\frac{\partial{U}}{\partial{\theta}}=R\,\sin{\alpha}\,F_{\theta},\,\,\,\nonumber \\
&& -\frac{\partial{U}}{\partial{\varphi}}=R\,\sin{\alpha}\,\sin{\theta}\,F_{\alpha}
\label{pot}
\end{eqnarray}
Now let us put Eqs.(\ref{kin}) and (\ref{pot}) into Lagrange equations
\begin{eqnarray}
\frac{d}{dt}\,\frac{\partial{T}}{\partial{v_{i}}}-
\frac{\partial{T}}{\partial{x_{i}}}=&&
\sum_{j=1}^{N_{c}} \frac{\partial{U_{ij}}}{\partial{x_{i}}} +
\sum_{k=1}^{N_{m}} \frac{\partial{U_{ki}}}{\partial{x_{i}}} 
\nonumber \\
&& + \sum_{l=1}^{2} \frac{\partial{U_{li}}}{\partial{x_{i}}}  \:\:\: 
(i=\alpha,\theta,\varphi \:\:\text{and}\:\: x_{i}=i)
\end{eqnarray}
where on the right hand side the first term arises from all counterions,
the second term is the macroionic attraction, the last term is due to
plane attraction. The Newton equations of counterion motion reads as follows
\begin{eqnarray}
\ddot{\alpha} =\sin{\alpha}\,\cos{\alpha}\,\left({\dot{\theta}}^{2} +
{\sin^{2}{\theta}} {\dot{\varphi}}^{2} \right) +\frac{1}{m\,R}\,F_{\alpha} , \, 
\label{alpha}
\end{eqnarray}

\begin{eqnarray}
\frac{d}{dt}\left({\sin^{2}{\alpha}}\,\dot{\theta}\right)=
{\sin^{2}{\alpha}}\,\sin{\theta}\,cos{\theta}\,{\dot{\varphi}}^{2} + 
\frac{\sin{\alpha}}{m\,R}\,F_{\theta} , 
\label{theta}
\end{eqnarray}

\begin{eqnarray}
\frac{d}{dt}\left({\sin^{2}{\alpha}}\, {\sin^{2}{\theta}}\,\dot{\varphi}\right)=
\frac{\sin{\alpha}\,\sin{\theta}}{m\,R}\,F_{\varphi}.
\label{fi}
\end{eqnarray}

We note that $(F_{\alpha},F_{\theta},F_{\varphi})$ are the components
of total force arising from all other counter-ions, planes and
macro-ions. The equations of motion Eq.(\ref{alpha})-(\ref{fi}) have been 
solved in a way similar to that described in~\cite{bajoras}.

The spherical components $(F_{\alpha},F_{\theta},F_{\varphi})$ of
vector $\vec{F}$ are connected with the 4D force eq.(~\ref{4dforce}) by the
following relations
\begin{eqnarray*}
F_{\alpha} = F_{1}\cos{\alpha}\sin{\theta}\cos{\varphi} &+&
 F_{2}\cos{\alpha}\sin{\theta}\sin{\varphi} \nonumber \\
&+&  F_{3}\cos{\alpha}\cos{\theta}-F_{4}\sin{\theta}, 
\end{eqnarray*}

\begin{eqnarray}
F_{\theta} = F_{1}\cos{\theta}\cos{\varphi} +
 F_{2}\cos{\theta}\sin{\varphi}-F_{3}\sin{\theta}, 
\end{eqnarray}

\begin{eqnarray*}
F_{\varphi} = -F_{1}\sin{\varphi} +F_{2}\cos{\varphi}. 
\end{eqnarray*} 

\bibliographystyle{unsrt}

%
\end{document}